\documentclass[notitlepage,twocolumn,amsmath,amssymb,aps,prl,superscriptaddress, 10pt]{revtex4-1}
\usepackage{graphicx}
\usepackage{dcolumn}
\usepackage{bm}
\usepackage[]{float}
\usepackage[]{hyperref}
\usepackage{verbatim}
\newcommand{\com}[1]{}
\usepackage{color}

\begin{document}

\title{Cavity-enhanced field-free molecular alignment at a high repetition rate}

\author{Craig Benko}
\email{craig.benko@colorado.edu}
\affiliation{JILA, NIST and the University of Colorado, 440 UCB, Boulder, CO 80309-0440, USA}
\author{Linqiang Hua}
\affiliation{JILA, NIST and the University of Colorado, 440 UCB, Boulder, CO 80309-0440, USA}
\affiliation{State Key Laboratory of Magnetic Resonance and Atomic and Molecular Physics, Wuhan Institute of Physics and Mathematics, Chinese Academy of Sciences, Wuhan 430071, China}
\author{Thomas K. Allison}
\affiliation{Departments of Chemistry and Physics, Stony Brook University, Stony Brook, New York 11794-3400, USA}
\author{Fran\c cois Labaye}
\affiliation{JILA, NIST and the University of Colorado, 440 UCB, Boulder, CO 80309-0440, USA}
\author{Jun Ye}
\email{ye@jila.colorado.edu}
\affiliation{JILA, NIST and the University of Colorado, 440 UCB, Boulder, CO 80309-0440, USA}

\begin{abstract}
\label{Abstract}Extreme ultraviolet frequency combs are a versatile tool with applications including precision measurement, strong-field physics, and solid-state physics. Here we report on an application of extreme ultraviolet frequency combs and their driving lasers to studying strong-field effects in molecular systems. We perform field-free molecular alignment and high-order harmonic generation with aligned molecules in a gas jet at 154 MHz repetition rate using a high-powered optical frequency comb inside a femtosecond enhancement cavity. The cavity-enhanced system provides means to reach suitable intensities to study field-free molecular alignment and enhance the observable effects of the molecule-field interaction. We observe modulations of the driving field, arising from the nature of impulsive stimulated Raman scattering responsible for coherent molecular rotations. We foresee impact of this work on the study of molecule-based strong-field physics, with improved precision and a more fundamental understanding of the interaction effects on both the field and molecules.
\end{abstract}
\maketitle

\label{Introduction}  Extreme ultraviolet (XUV) frequency combs are capable of having spatial and temporal coherence properties that rival visible light~\cite{Benko:2014} and are among the brightest sources of XUV radiation originating from HHG, now approaching the average brightness of syncrotrons~\cite{Yost:2011b,Lee:2011,Allison:w2011}. Since their initial development \cite{Jones:2005,Gohle:2005}, XUV frequency combs have had found several applications including precision spectroscopy~\cite{Cingoz:2012}, strong-field physics~\cite{Yost:2009,Benko:2014}, and solid-state physics~\cite{Mills:2014}. Recently, their spectral extent reached wavelengths as short as 10 nm~\cite{Pupeza:2013}. The main tool for generating XUV frequency combs has been the combination of optical frequency combs and passive enhancement cavities that are capable of supporting femtosecond pulse bandwidths, the so-called femtosecond enhancement cavities (fsECs)~\cite{Jones:2002}.

Reaching intensities suitable for HHG at high repetition rate also opens the door to study a variety of other nonlinear optical phenomena. Ultrashort laser pulses are capable of aligning molecules by an impulsive stimulated Raman scattering (ISRS) process. In the impulsive limit, when the excitation pulse duration is shorter than the characteristic rotational period, coherent rotations persist long after the excitation pulse has passed through the molecular sample. The resulting revival structure can be exploited for performing experiments on molecules aligned in the lab frame in the absence of control fields, or so-called field-free molecular alignment (FFMA)\cite{Seideman:1995,Ortigoso:1999}. This has been exploited in pioneering experiments performing dynamic imaging of molecular structure \cite{Itatani:2004, Hockett:2011}. Of particular interest to strong-field physics is the interaction of molecules with intense laser fields~\cite{Seideman:1995,Stapelfeldt:2003} and performing HHG with aligned molecules~\cite{Velotta:2001,Itatani:2004,Itatani:2005,Mairesse2008a,Salieres:2012}.

XUV comb systems potentially offer large factors of improvement in the fidelity of these studies due to their high signal to noise \cite{Yost:2009} and coherence properties \cite{Benko:2014}. In this Letter we study molecular alignment and HHG from aligned molecules at 154 MHz repetition rate using a fsEC system. This represents a nearly 5 order-of-magnitude increase in the the repetition rate compared to typical experiments performing FFMA-based strong-field studies~\cite{Itatani:2004,Itatani:2005,Lock:2012}. With the molecular sample rotationally excited inside our enhancement cavity, we investigate the effects of ISRS on the driving laser field by analyzing the cavity-transmitted light of a second, time-delayed pulse. We also perform HHG with the aligned molecules using the second pulse. The FFMA-based HHG at high repetition rate will facilitate heterodyne interferometry of the HHG signal in the extreme ultraviolet, permitting precise access to both amplitude and phase information of the XUV light~\cite{Benko:2014} as a function of molecular alignment.

The impulsive nature of ISRS can have important effects on the driving laser~\cite{Yan:1985,Nazarkin:1998,Korn:1998}. The laser pulse experiences a self-phase modulation-like process, which leads to a red-shifting of the original optical spectrum. The process occurs without an intensity threshold. These features are distinct from SRS in the adiabatic regime~\cite{Yan:1985,Nazarkin:1998,Korn:1998}. Impulsive FFMA can also have important consequences for precision studies of strong-field physics~\cite{Seideman:1995,Velotta:2001,Itatani:2004,Mairesse2008a,Stapelfeldt:2003,Salieres:2012,Itatani:2005} in aligned molecules since the HHG-driving laser pulse will experience spectral and phase shifts related to ISRS during the HHG process. Despite FFMA being a well understood phenomenon~\cite{Seideman:1995,Stapelfeldt:2003}, the effects on the driving laser originating from ISRS are often ignored in HHG and FFMA experiments due to their small effects when thin, freely expanding gas targets are used at low densities. However, in the cavity-enhanced approached describe in this work, we are more sensitive to the effects of ISRS on the driving laser due to the cavity effectively increasing the interaction length of our sample by a factor proportional to the cavity finesse. Indeed, we can now provide a clean measurement of the effects of ISRS with our enhancement cavity approach while maintaining otherwise similar experimental conditions to conventional molecular HHG experiments. Our measurements indicate that modulations (amplitude and phase) of the driving laser cannot be ignored in future experiments utilizing fsECs and must be characterized to properly describe the XUV frequency comb spectral amplitude and phase.

\begin{figure}[]
\includegraphics[scale = .39]{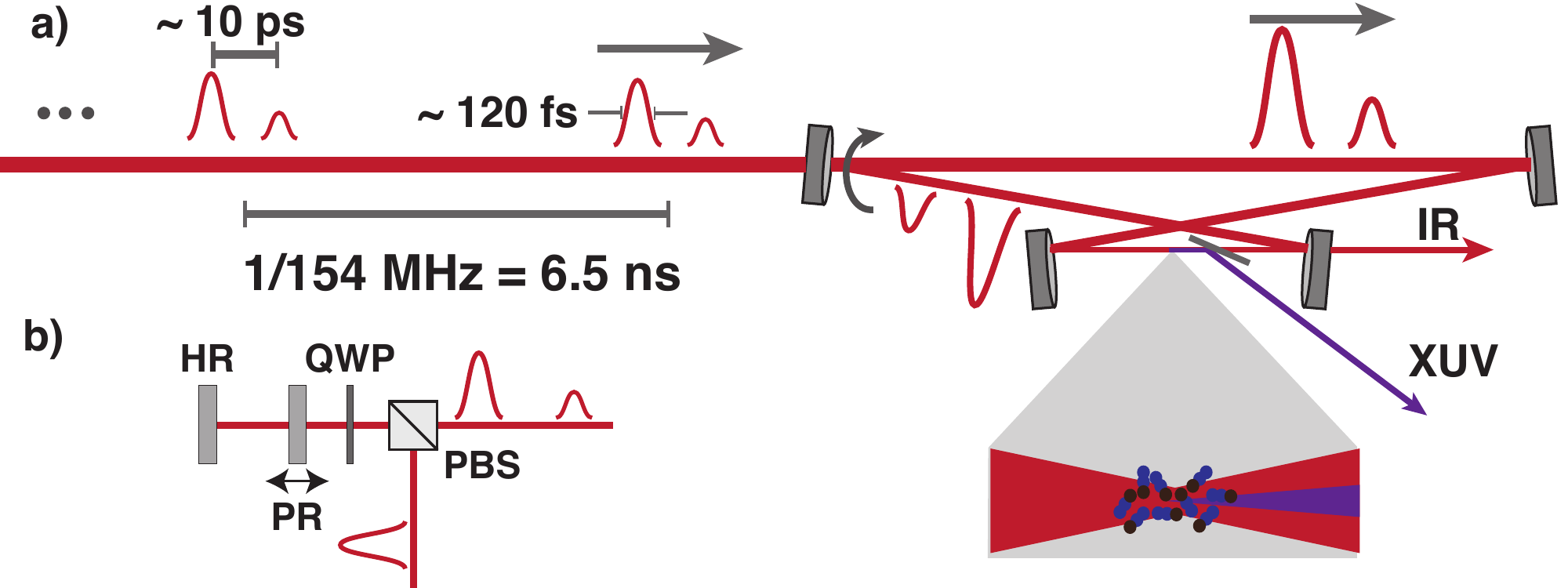}
\caption{\label{fig:Fig1} {\bf a)} Schematic of the experiment.  A high repetition-rate train of dual pulses, one for alignment and the second for HHG, are coherently coupled into a high-finesse, femtosecond enhancement cavity. At the focus of the cavity, the molecules are aligned by the first pulse and high-order harmonic generation is performed by the second pulse. The XUV light can be extracted from the cavity using a sapphire plate placed at Brewster's angle for the driving laser. {\bf b)} The dual pulse feature of femotsecond pulse train is produced from a Gires-Tournois interferometer, with a tunable delay between the two pulses. HR, high reflector. PR, partial reflector. QWP, quarter waveplate. PBS, polarizing beamsplitter.}
\end{figure}

\label{Experimental Details} In our experiment, we use an 80 W, 120 fs, 154 MHz repetition rate Yb:fiber frequency comb centered at 1070 nm to drive a fsEC~\cite{Ruehl:2010}. The resonance modes of the cavity are locked to the corresponding lines of the incident optical frequency comb. A piezoelectric transducer on a cavity mirror actuates on error signals derived from radio frequency sidebands on the frequency comb~\cite{Jones:2004}. The cavity can operate with a power enhancement factor ranging from 200$-$400 and can achieve up to 10 kW of average power. For this experiment,  10 kW of average power corresponds to a pulse peak intensity of \mbox{$1\times10^{14}$ W cm$^{-2}$} at the focus of the cavity at the full repetition rate. Accurate intensity calibration is achieved by the well defined optical mode inside the enhancement cavity and measurement of the enhancement cavity parameters~\cite{Jones:2002,Allison:w2011}. The intracavity pulse duration was verified with second-harmonic generation intensity autocorrelation. The intracavity spectrum is measured with an optical spectrum analyzer using the light transmitted through a cavity high-reflecting mirror.

The apparatus is schematically shown in Fig.~\ref{fig:Fig1}. Performing HHG with aligned molecules requires two successive pulses, the first to rotationally excite the molecules and the second to drive HHG (henceforth, pump and probe respectively). To accomplish this goal, we use a mirror combination similar to a Gires-Tournois interferometer (GTI) to convert the original laser pulse train into two with a tunable timing delay, schematically shown in Fig.~\ref{fig:Fig1} b). The combination of a partial reflector \mbox{(R $\approx .1$)} and a high reflector \mbox{(R $\approx 1$)} along with a quarter \mbox{($\lambda/4$)} waveplate and a polarizing beamsplitter cube generates the dual-pulse train. This combination of optics also generates weaker pulses after the probe pulse, but these do not effect the rotational dynamics generated by the pump and observed by the much larger probe. Additionally, the GTI combination of optics is more efficient in generating the pulse train than a more traditional Michelson-type interferometer that relies on polarization optics. By tuning the value of the partial reflector, the relative heights of the pump and probe can be adjusted. By simply changing the distance between the partial and high reflector, the timing delay between the pulses can be tuned. The pulse train after the GTI combination was analyzed with second-harmonic intensity autocorrelation to ensure the pulse separation and ultrashort pulse durations were maintained. For our experiment, the intensity ratio of alignment pulse (pump) to HHG-driving pulse (probe) is 1:8; however, this value can be varied from 1:3 to 1:30 only limited by the available PR mirrors. The pulse train is coherently coupled into the fsEC, and at the cavity focus we inject N$_2$O gas through a quartz nozzle with a \mbox{$\sim$~120 $\mu$m}
diameter with up to $\sim$~3 atmospheres (atm) of constant backing pressure. XUV light produced with HHG is outcoupled from the cavity using a 250 $\mu$m thick sapphire plate placed at Brewster's angle for the fundamental driving laser.

The theory of FFMA of linear molecules is well established and we refer to Ref.~\cite{Seideman:1995,Ortigoso:1999} for a detailed discussion. To summarize, the molecules receive an impulsive kick from the laser pulse, and many rotational states are coupled owing to the large bandwidth associated with the ultrashort pulse. After the pulse has passed, the molecular wave function then evolves freely. At the rotational period $T_r$ and its integer multiples, the molecules will exhibit strong alignment and anti-alignment (revivals) along the polarization axis of the laser. Depending on the molecule under investigation, revivals also occur at fractions of $T_r$. The degree of alignment is quantified by the thermally averaged expectation value $\langle\langle\text{cos}^2\theta\rangle\rangle$.

\begin{figure}[]
\includegraphics[scale = .33]{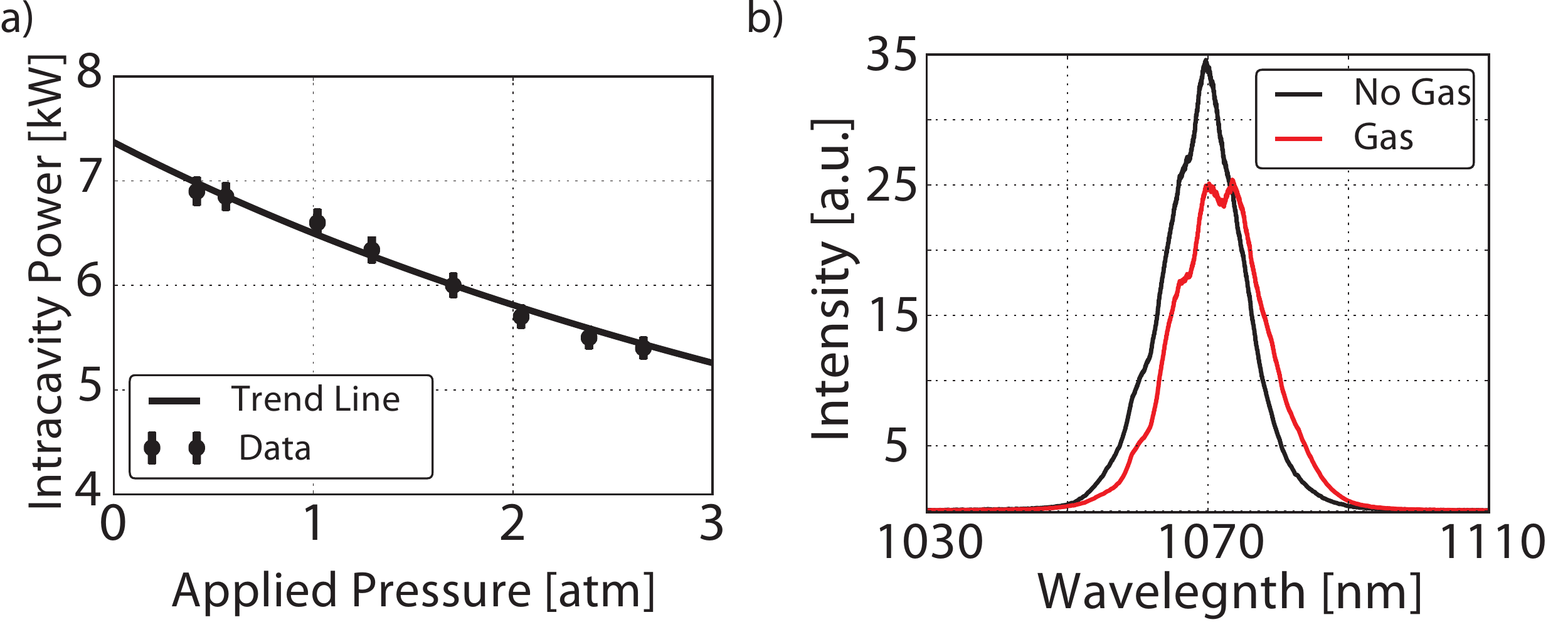}
\caption{\label{fig:Fig2} {\bf a)} Pressure dependence of the buildup power when a single pulse is interacting with molecules inside the cavity at $f_{rep}$ = 154 MHz.  As the N$_2$O pressure is increased, a decrease in the intracavity power is observed due to loss of power enhancement in the cavity. A trend line is shown with the data. {\bf b)} The intracavity spectrum exhibits a clear shift to the red when gas is present, a signature of impulsive stimulated Raman scattering. The red shift is dominant even in the presence of ionization, which would shift the spectrum to the blue. The red shifted spectra corresponds to the data in (a) at 1.7 atm of applied pressure.}
\end{figure}

\label{Data Single Pulse} Our work begins with investigating the effects of a single pulse propagating and interacting with an N$_2$O gas jet inside the fsEC. N$_2$O was chosen because of the relatively large anisotropic polarizability to facilitate alignment, the small rotational constant to create a large delay between revivals, and the low ionization potential to facilitate observation of above-threshold harmonics. We estimate the rotational temperature of the N$_2$O gas to be 30-60 K~\cite{Miller1988}. The pulse has a peak intensity of \mbox{$0.7 \times 10^{14}$ W cm$^{-2}$} in the absence of molecules. We monitor both the intracavity spectrum and power as a function of the applied pressure to the gas nozzle. As shown in Fig.~\ref{fig:Fig2}, we observe a clear red shift of the spectrum, accompanied with a systematic decrease of the power. The power decrease results from spectral red-shifting decreasing the enhancement of the cavity and not from light scattering out of the cavity. The red shift is contrary to what is usually observed \mbox{-- spectral blue shift --} when a femtosecond pulse propagates through ionizing media~\cite{Allison:w2011}.  This demonstrates that ISRS has a more dominant effect on the pulse when interacting with N$_2$O at these intensities and densities. The red shift is a manifestation of energy transfer from the field to the molecular rotations.

\begin{figure}[]
\includegraphics[scale = .31]{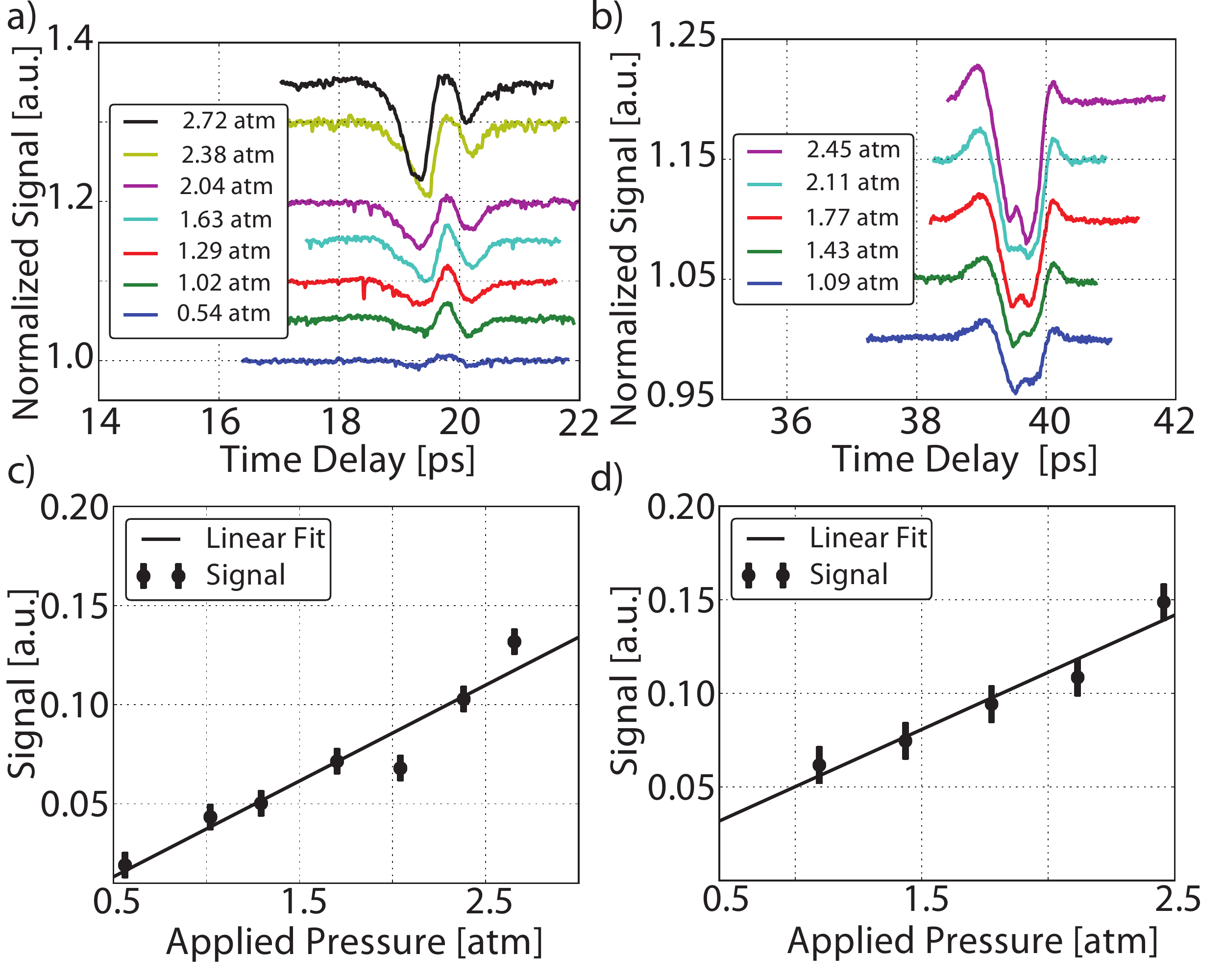}
\caption{\label{fig:Fig3} {\bf a,b)}  The effects of the half-revival (a) and full-revival (b) on the probe pulse as a function of pressure on the intracavity power. The signal is normalized to the baseline ($\sim 16$ ps and $\sim 36$ ps, respectively) and offset vertically for clarity. The pressure reflects what is applied to the 120 $\mu$m diameter quartz nozzle. {\bf c,d)} The size of the half (c) and full (d) revival signals, determined with $|$max(signal)-min(signal)$|$ of the revival structure. The dependence of the signal size on sample density is fit to a linear function.}
\end{figure}

\label{Data Multi pulse}We proceed to inject two pulse trains into the fsEC. The first pulse (pump) excites a rotational wave-packet. The second pulse (probe) is of greater intensity and generates high-order harmonics. We observe two important features as a function of delay between the two pulses. First, the HHG yield of the probe pulse is modulated at the revivals of N$_2$O (more on this later). Second, thanks to the cavity-based measurement and the multi-pass effect, we observe clear effects of the pump pulse on the probe pulse that occur at the rotational revivals of N$_2$O. We observe these effects by monitoring the intracavity spectrum and power relying on the intracavity spectrum is dominated by the much more intense probe pulse and that the delay-dependent effects can only affect the probe. The effects on the probe pulse persist even if the probe intensity is too weak for appreciable \mbox{HHG ( $< 0.2 \times 10^{14}$ W cm$^{-2}$)}.

\begin{figure}[]
\includegraphics[scale = .38]{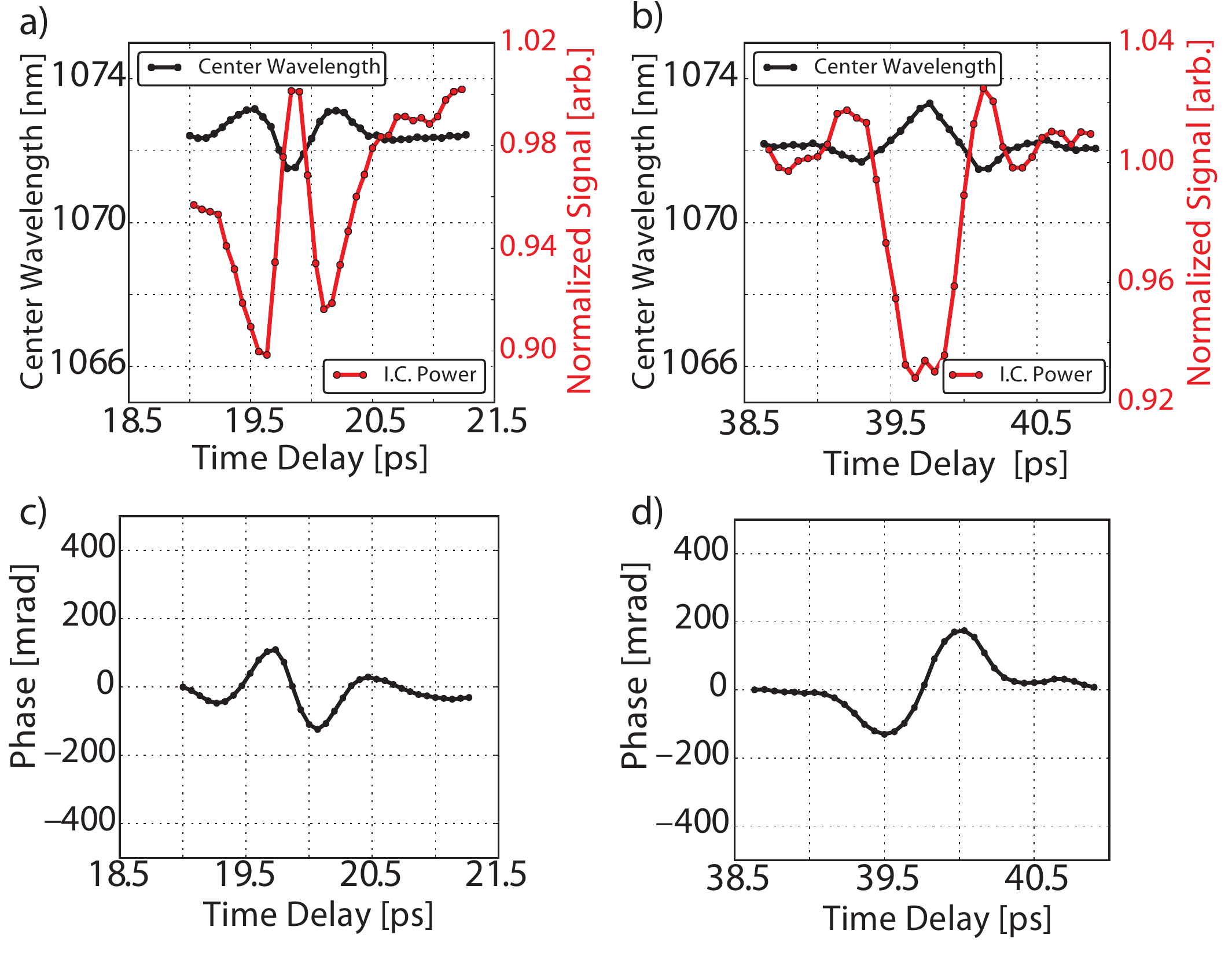}
\caption{\label{fig:Fig4} {\bf a,b)} The intracavity central wavelength and the intracavity power are shown as a function of pump-probe delay for the half-revival (a) and full-revival (b) of N$_2$O. {\bf c,d)} The phase as a function of delay for the half-revival (c) and full-revival (d) of N$_2$O respectively using the relation $\omega(t) = \omega_0 + d\phi/dt$. The conditions for the data were the same as Fig.~\ref{fig:Fig3}a,b with 2 atm of pressure for the half-revival and full-revival data.}
\end{figure}

The effects on intracavity power are presented in Fig.~\ref{fig:Fig3}. At $\sim$20 ps and $\sim$40 ps of delay between the pump and probe, the half- and full-revival of N$_2$O alignment, respectively, modulate the intracavity power, as shown in Fig.~\ref{fig:Fig3} a,b. The intensity uncertainty is $\sim 1 \%$. This power modulation is related to spectral shifts of the intracavity spectrum, shown in Fig.~\ref{fig:Fig4}a,b as a function of pump-probe delay for the half- and full-revival respectively. The dominant contribution to the modulations in the power arises from the reduced overlap between the intracavity comb and the exciting laser. The  center of the intracavity spectrum shifts as the delay is scanned along with the corresponding intracavity power as shown in Fig.~\ref{fig:Fig4}a,b on the left  vertical axis. The conditions for the half-revival and full-revival data were the same as Fig.~\ref{fig:Fig3}a,b at 2 atm of nozzle backing pressure. The increase in intracavity power occurs with a slight blue-shift in the spectrum and the decrease with a slight red-shift. When the laser spectrum shifts to the blue, the spectrum becomes closer to the empty cavity case. The improved overlap leads to an increase of the intracavity power. Conversely, a red-shift reduces the spectral overlap further and the intracavity power decreases. It is important to note that in the single pulse case (e.g., Fig.~\ref{fig:Fig2}), only red shift of the spectrum is observed. However, with the combination of pump and probe pulses, the molecular alignment/coherent rotation can transfer energy back and forth between the molecule and the field, depending on the pump-probe delay time~\cite{Yan:1985}.

\label{Systematic scalings}To provide a systematic investigation of the intracavity response, we have measured its dependence on a number of important parameters including the target gas density and pump/probe intensities. As the molecular density is increased, we observe a linear increase in the size of the revival effect on the intracavity power (measured by \mbox{$|$max(signal)-min(signal)$|$} of the revival structure), as shown in Fig.~\ref{fig:Fig3}c,d. This dependence is consistent with ISRS~\cite{Yan:1985}. We also see a linearly increasing response as either the pump or probe is increased independently. This effect is also consistent with ISRS and the degree of alignment $\langle\langle\text{cos}^2\theta\rangle\rangle $ increasing linearly in the small intensity limit. These observations have been verified by numerically simulating the alignment process.

\label{Phase}To complete our investigation of the field-molecule interaction, we use the observed spectral shifts to estimate the effects of ISRS on the phase of the probe pulses. We determine the phase shifts by analyzing the measured central wavelength of the intracavity spectrum as a function of pump-probe delay. The delay dependent spectral shifts are smaller than the red-shift observed when the gas is introduced (see Fig.~\ref{fig:Fig2}b.). The spectral shifts and their corresponding intracavity power are shown in Fig.~\ref{fig:Fig4}a,b for the half- and full-revival respectively. Using the relation $\omega(t) = \omega_0 + d\phi/dt$, we extract the phase dependence as a function of delay \cite{Bartels:2001}. The phase shift results are shown in Fig.~\ref{fig:Fig4}c,d for half- and full-revival respectively. Large phase shifts are introduced to the driving laser. This will be important to understand and control for experiments with HHG because small phase shifts on the pump will be transferred to the harmonic light and scale with harmonic order~\cite{Benko:2014}.

\begin{figure}[]
\includegraphics[scale = .34]{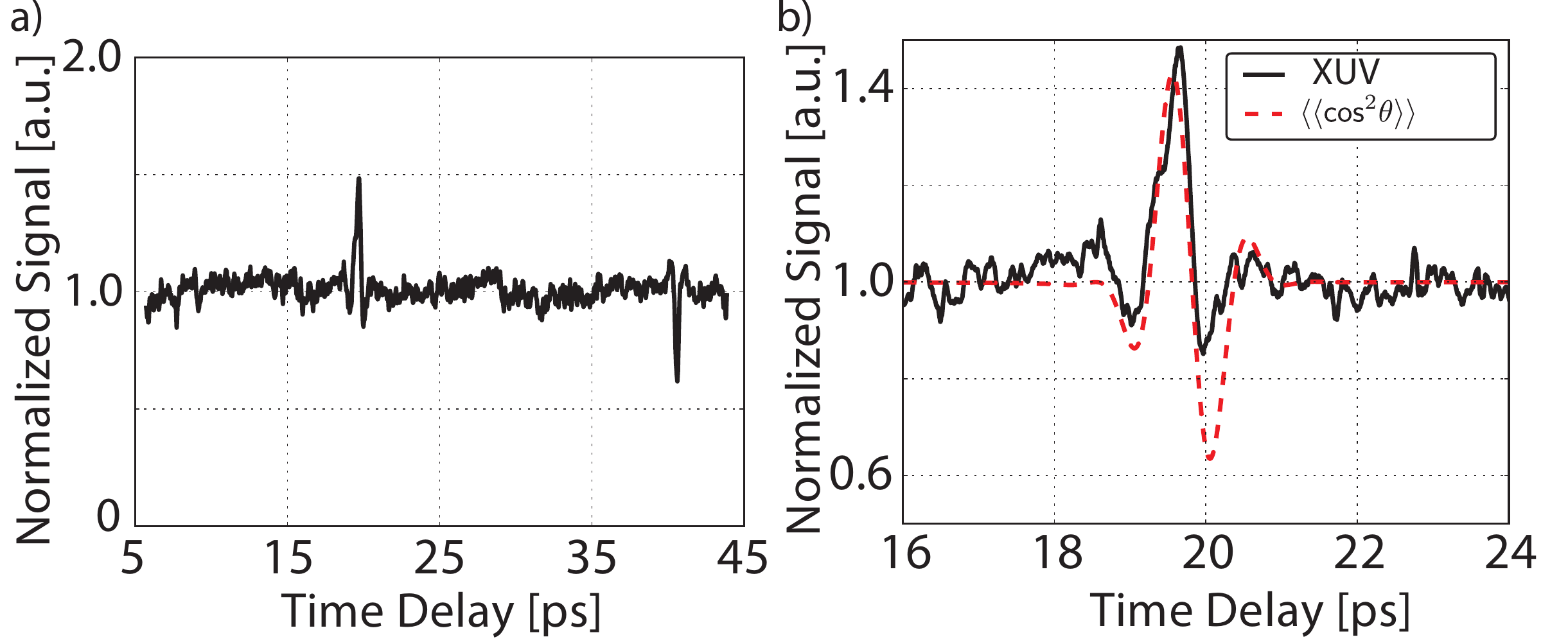}
\caption{\label{fig:Fig5} {\bf a)} High harmonic signal generated from aligned N$_2$O molecules.  Harmonics 15-19 are measured as a function of pump-probe delay.  The half and full revival of N$_2$O is shown at $\sim$19 and $\sim$ 39 ps of delay, respectively. Revival signals on the harmonic yield are observed out to $\sim$ 80 ps without a large decay of signal. {\bf b)} The amplitude modulation on the XUV yield is shown in detail. The $\langle\langle\text{cos}^2\theta\rangle\rangle$ expectation value is shown for reference.}
\end{figure}

\label{XUV Data} With this clear determination of the molecular alignment effect on the pulse that is used to drive HHG, we now turn our attention to the measurement of the HHG yield as a function of molecular alignment. A pump pulse of \mbox{$7.5 \times 10^{12}$ W cm$^{-2}$} is used to prepare a rotational wavepacket before a probe pulse of \mbox{$0.6 \times 10^{14}$ W cm$^{-2}$} is used to perform HHG. The total yield of harmonics 15 - 19 are detected simultaneously with an electron multiplier as the delay between the pump and probe pulses is scanned. These harmonics were isolated with an aluminum filter and the bandwidth of our B$_4$C optics. The data is documented in Fig.~\ref{fig:Fig5}. The XUV yield is normalized to the baseline at \mbox{t $\sim$ 5 ps}. The delay was scanned continuously at a rate of at \mbox{1 ps/s} and no averaging of the data was performed besides low-pass filtering at 300 Hz. Further averaging could be performed to improve signal to noise and is not presented here. In Fig.~\ref{fig:Fig5}b, we report the observation of a nearly
50\% modulation in the XUV yield. The XUV yield is shown with the $\langle\langle\text{cos}^2\theta\rangle\rangle$ expectation value for reference. The similarity between the data and $\langle\langle\text{cos}^2\theta\rangle\rangle$ is expected when the harmonics are far from an interference region~\cite{Lein2002, Vozzi2005}. This XUV yield does not mimic the effect on the driving laser, meaning that the modulations in the XUV yield are not driven by modulations in the driving probe power. With our fsEC apparatus we are easily able to achieve probe intensities of $1 \times 10^{14}$ W cm$^{-2}$ and we have also performed HHG experiments with molecules of higher ionization threshold such as CO$_2$ or harder to ionize molecules like O$_2$ and observed similar modulations in the HHG yield arising from rotational revivals.

\label{Conclusion} In conclusion, we have demonstrated FFMA and molecular HHG at high repetition rate. We have made precise investigations of the interaction between laser and molecules by generating alignment and observing the effect of ISRS on both the driving laser field and the molecules, thus providing a complete picture of the interaction. Observing HHG from aligned molecules at high repetition rates opens exciting avenues for probing molecular rotational dynamics~\cite{Lock:2012,Itatani:2005} with more rapid data acquisition and increased measurement precision. It may also be possible to enhance the degree of alignment with repetitive impulsive kicks~\cite{Cryan:2009}. With the recent demonstration of heterodyne interferometry with XUV combs~\cite{Benko:2014} we expect to measure both the amplitude and phase of XUV light from aligned molecules~\cite{Itatani:2004}.

\label{Acknowledgement}
\begin{acknowledgments}This work is supported by the National Institute of Standards and Technology, Air Force Office of Scientific Research (AFOSR), the DARPA PULSE program, and the National Science Foundation Physics Frontier Center at JILA. TKA acknowledges support from AFOSR Young Investigator Research Program FA9550-13-1-0109.
\end{acknowledgments}

\bibliographystyle{apsrev4-1}
\bibliography{ISRS}
\end{document}